\documentclass[twocolumn,showpacs,preprintnumbers,amsmath,amssymb]{revtex4}

\usepackage{natbib}
\usepackage{lgrind}
\usepackage{latexcad}
\usepackage{float}
\usepackage{amssymb}
\usepackage{amsmath,amsfonts}
\usepackage{graphicx}
\usepackage{rotating}
\usepackage{dcolumn}
\usepackage{mathrsfs}
\usepackage{bm}

\begin{document}


\title{Time-Periodic Solutions of the Einstein's Field Equations}
\author{De-Xing Kong$^1$ and Kefeng Liu$^{2,1}$} \affiliation{\\ $^{1}$\!Center of Mathematical Sciences,
Zhejiang University, Hangzhou 310027, China\\
$^2$\!Department of Mathematics, University of California at Los
Angeles, CA 90095, USA}
\date{\today}

\begin{abstract}
In this paper, we develop a new method to find the exact solutions
of the Einstein's field equations by using which we construct
time-periodic solutions. The singularities of the time-periodic
solutions are investigated and some new physical phenomena, such as
the time-periodic event horizon, are found. The applications of
these solutions in modern cosmology and general relativity are
expected.
\end{abstract}

\pacs{04.20.Jb; 04.20.Dw; 98.80.Jk; 02.30.Jr}

\keywords{Einstein's field equations, time-periodic solutions, event
horizon.}

\maketitle

{\em 1. Introduction.} The Einstein's field equations are the
fundamental equations in general relativity and cosmology. The
general version of the gravitational field equations or the
Einstein's field equations read
\begin{equation}
R_{\mu\nu}-\frac{1}{2}g_{\mu\nu}R+\Lambda g_{\mu\nu}=\frac{8\pi
G}{c^4}T_{\mu\nu},
\end{equation}
where $g_{\mu\nu}\;(\mu,\nu=0,1,2,3)$ is the unknown Lorentzian
metric, $R_{\mu\nu}$ is the Ricci curvature tensor,
$R=g^{\mu\nu}R_{\mu\nu}$ is the scalar curvature, where $g^{\mu\nu}$
is the inverse of $g_{\mu\nu}$, $\Lambda$ is the cosmological
constant, $G$ stands for the Newton's gravitational constant, $c$ is
the velocity of the light and $T_{\mu\nu}$ is the energy-momentum
tensor. In a vacuum, i.e., in regions of space-time in which
$T_{\mu\nu}=0$, the Einstein's field equations (1) reduce to
\begin{equation}
R_{\mu\nu}-\frac{1}{2}g_{\mu\nu}R+\Lambda g_{\mu\nu}=0,
\end{equation}
or equivalently,
\begin{equation}
R_{\mu\nu}=\Lambda g_{\mu\nu}.
\end{equation}
In particular, if the cosmological constant $\Lambda$ vanishes,
i.e., $\Lambda =0$, then the equation (2) becomes
\begin{equation}
R_{\mu\nu}-\frac{1}{2}g_{\mu\nu}R=0,
\end{equation}
or equivalently,
\begin{equation}
R_{\mu\nu}=0.
\end{equation}
Each of the equations (2)-(5) can be called the vacuum Einstein's
field equations.

The mathematical study on the Einstein's field equations includes,
roughly speaking, the following two aspects: (i) establishing the
well-posedness theory of solutions; (ii) finding exact solutions
with physical background. Up to now, very few results on the
well-posedness for the Einstein's field equations have been
established. In their classical monograph \cite{ck}, Christodoulou
and Klainerman proved the global nonlinear stability of the
Minkowski space for the vacuum Einstein's field equations, i.e. they
showed the nonlinear stability of the trivial solution of the vacuum
Einstein's field equations. Recently, by using wave coordinates,
Lindblad and Rodnianski gave a simpler proof (see \cite{lr}). In her
Ph.D. thesis \cite{z}, Zipser generalizes the result of
Christodoulou and Klainerman \cite{ck} to the Einstein-Maxwell
equations. As for finding exact solutions, many works have been done
and many interesting results have been obtained (see, e.g.,
\cite{c}, \cite{skmhh}, \cite{b}). In what follows, we will briefly
recall some basic facts about the exact solutions of the Einstein's
field equations.

The exact solutions are very helpful to understand the theory of
general relativity and the universe. The typical examples are the
Schwarzschild solution and the Kerr solution (see \cite{c}). These
solutions provide two important physical space-times: the
Schwarzschild solution describes a stationary, spherically symmetric
and asymptotically flat space-time, while the Kerr solution provides
a stationary, axisymmetric and asymptotically flat space-time.

The study on exact solutions of the Einstein's field equations has a
long history. In December 1915, Karl Schwarzschild discovered the
first non-trivial solution to the vacuum Einstein's field equations
which is a static solution with zero angular momentum (see
\cite{sch}). The unique two-parameter family of solutions which
describes the space-time around black holes is the Kerr family
discovered by Roy Partrick Kerr in July 1963 (see \cite{kerr}).
These solutions are very important in studying black holes in the
Nature which is just the study of these solutions (see \cite{c}).
Various generalizations of the Kerr solution have been done (see,
e.g., \cite{skmhh} and \cite{b}). Gowdy \cite{gowdy1}-\cite{gowdy2}
constructed a new kind of solutions of the vacuum Einstein's
equations, these solutions provide a new type of cosmological model.
This model describes a closed inhomogeneous universe, space sections
of these universes have either the three-sphere topology $S^3$ or
the wormhole (hypertorus) topology $S^1\otimes S^2$. Recently, Ori
\cite{o} presented a class of curved-spacetime vacuum solutions
which develop closed timelike curves at some particular moment, and
used these vacuum solutions to construct a time-machine model. The
Ori model is regular, asymptotically flat, and topologically
trivial.

From the above discussions, we see that the exact solutions play a
crucial role in general relativity and cosmology, so it is always
interesting to find new exact solutions for the Einstein's
equations. Although many interesting and important solutions have
been obtained, there are still many fundamental but open problems.
One interesting open problem is {\it if there exists a
``time-periodic" solution to the Einstein's field equations}. One of
the main results in this paper is a solution of this problem.

In this paper, we focus on finding the exact solutions of the vacuum
Einstein's field equations (3) and (5). We will present a new method
to find exact solutions. Using this method we can construct
interesting and important exact solutions, for example, the
time-periodic solution of the vacuum Einstein's field equations. We
analyze the singularities of time-periodic solutions and investigate
some new physical phenomena enjoyed by these new space-times. We
find that the new time-periodic solutions have time-periodic event
horizon, which is a new phenomenon in the space-time geometry. The
applications of these solutions and their new properties in modern
cosmology and general relativity may be expected.

More precisely, our time-periodic solution to the vacuum Einstein's
field equations in the spherical coordinates $(t,r,\theta,\varphi)$
can be written in the following form
\begin{equation} \label{eq:4.6}
ds^2=(dt,dr,d\theta,d\varphi)(\eta_{\mu\nu})(dt,dr,d\theta,d\varphi)^T,
\end{equation}
where
\begin{equation} \label{eq:4.7}
(\eta_{\mu\nu})= \left(
\begin{array}{cccc}
G & -G+\displaystyle\frac{Mr}{r-m} & QK & 0  \\
-G+\displaystyle\frac{Mr}{r-m} & G-\displaystyle\frac{2Mr}{r-m} & -QK & 0  \\
QK & -QK & -K^2 & 0 \\
0 & 0 & 0 & -K^2\sin^2\theta \\
\end{array}\right),
\end{equation}
in which
\begin{equation} \label{eq:4.8}
\left\{\begin{array}{l}
G=1+2\varepsilon \Omega^+\sin\theta\cos{(t-r)},\\
K=r+m\ln{|r-m|}+\varepsilon\sin{(t-r)},\\
M=\Omega^+\sin\theta,\\
Q=-\displaystyle\frac{1}{2}(1+2\sin\theta)\Omega^-.
\end{array}\right.
\end{equation}
In the above, $\varepsilon\in (-\frac 18,\frac 18)$ and $m\in
\mathbb{R}$ are two parameters, and $\Omega^{\pm}$ are defined by
\begin{equation} \label{eq:4.4}
\Omega^{\pm}=|\tan\theta/2|^{\frac12}\pm|\tan\theta/2|^{-\frac12}.
\end{equation}
In Section 5, we analyze the singularity behaviors and find some new
physical phenomena.

According to the authors' knowledge, (6) gives the first
time-periodic solution to the Einstein's field equations. Here we
would like to point out that, by using our method, we can re-derive
almost all known exact solutions to the vacuum Einstein's field
equations, for examples, G\"{o}del's solution \cite{go},
Khan-Penrose's solution \cite{kp}, Gowdy's solution
\cite{gowdy1}-\cite{gowdy2}, etc. Our method can also be used to
find exact solutions of the Einstein's field equations in higher
dimensions which will be of interests in string theory.

{\em 2. Lorentzian metrics.} The Einstein's field equations are a
second order global hyperbolic system of highly nonlinear partial
differential equations with respect to the Lorentzian metric
$g_{\mu\nu}\;(\mu,\nu=0,1,2,3)$. To solve the Einstein's field
equations, one key point is to choose a suitable coordinate system.
A good coordinate system can simplify the equations and make them
easier to solve. In the study on the Einstein's field equations,
there are three famous coordinate systems: harmonic coordinates,
wave coordinates and the Gaussian coordinates. In this paper, we
consider the metric of the following form
\begin{equation}\left(g_{\mu\nu}\right)=\left( \begin{array}{cccc}
g_{00} & g_{01} & g_{02} & g_{03}\\ g_{10} & g_{11} & 0 & 0 \\
g_{20} & 0 & g_{22} & 0\\
g_{30} & 0 & 0 & g_{33}\\
\end{array}\right),\end{equation}
where $g_{\mu\nu}$ are smooth functions of the coordinates
$(x^0,x^1,x^2,x^3)$ and satisfy $g_{0i}=g_{i0}$. In the coordinates
$(x^0,x^1,x^2,x^3)$, the line element reads $
ds^2=g_{\mu\nu}dx^{\mu}dx^{\nu}$. For simplicity of notations, we
denote the coordinates $(x^0,x^1,x^2,x^3)$ by $(t,x,y,z)$ and
rewrite (10) as
\begin{equation}\left(g_{\mu\nu}\right)=\left( \begin{array}{cccc}
u & v & p & q \\
v & w & 0 & 0 \\
p & 0 & \rho & 0 \\
q & 0 & 0 & \sigma \\
\end{array}\right),\end{equation}
where $u,v,p,q,w,\rho$ and $\sigma$ are smooth functions of the
coordinates $(t,x,y,z)$. It is easy to verify that the determinant
of $(g_{\mu\nu})$ is given by
\begin{equation}g\stackrel{\triangle}{=}\det(g_{\mu\nu})=
uw\rho\,\sigma-{v}^{2}\rho\,\sigma-{p}^{2}w\sigma-{q}^{2}w\rho\end{equation}
Throughout this paper, we assume that
$$g<0\eqno{(H)}$$
On the other hand, it is easy to see that at least two of the
functions $w,\rho,\sigma$ have the same sign. Without loss of
generality, we may suppose that $\rho$ and $\sigma$ keep the same
sign, for example,
\begin{equation}
\rho < 0\quad\rm{and} \quad\sigma < 0.\end{equation}

We can easily prove the following theorem.

\noindent{\bf Theorem 1} {\em Under the assumptions (H) and (13),
the metric $(g_{\mu\nu})$ is Lorentzian.}

We now consider the solutions of the Einstein's field equations. By
Bianchi identities, we believe that the general form of the
solutions of the Einstein's field equations takes one of the
following forms
$$\left(\eta_{\mu\nu}\right)\stackrel{\triangle}{=}\left(
\begin{array}{cccc}
u & v & p & q \\
v & 0 & 0 & 0 \\
p & 0 & \rho & 0 \\
q & 0 & 0 & \sigma \\
\end{array}\right),\eqno{(\rm{Type\; I})}$$
$$\left(\eta_{\mu\nu}\right)\stackrel{\triangle}{=}\left(
\begin{array}{cccc}
0 & v & p & q \\
v & w & 0 & 0 \\
p & 0 & \rho & 0 \\
q & 0 & 0 & \sigma \\
\end{array}\right)\eqno{(\rm{Type\; I\!I})}$$
or
$$\left(\eta_{\mu\nu}\right)\stackrel{\triangle}{=}\left(
\begin{array}{cccc}
u & v & p & 0 \\
v & w & 0 & 0 \\
p & 0 & \rho & 0 \\
0 & 0 & 0 & \sigma \\
\end{array}\right).\eqno{(\rm{Type\; I\!I\!I})}$$

For type I, the assumption (H) is equivalent to $ v\neq 0.$
Therefore, by Theorem 1 we have

\noindent{\bf Conclusion 1} {\em If
\begin{equation}
 \rho<0,\quad \sigma <0 \quad{\rm{and}} \quad v\neq 0,
\end{equation}
then the metric $(\eta_{\mu\nu})$ is Lorentzian.}

For type I$\!$I, we have

\noindent{\bf Conclusion 2} {\em If $w,\rho,\sigma$ are all negative
functions and $v^2+p^2+q^2\neq 0$, then the hypotheses (H) and (13)
are satisfied, and the metric $(\eta_{\mu\nu})$ is Lorentzian.}

Similarly, for type I$\!$I$\!$I we have

\noindent{\bf Conclusion 3} {\em If $u$ is positive and
$w,\rho,\sigma$ are negative, then the hypotheses (H) and (13) are
satisfied, and the metric $(\eta_{\mu\nu})$ is Lorentzian.}

We are interested in finding exact solutions of the Einstein's field
equations of the above types I-I\!I\!I.

{\em 3. New method to find exact solutions.} In order to illustrate
our method, as an example, we use the Lorentzian metric type I to
construct some interesting exact solutions, in particular
time-periodic solution for the vacuum Einstein's field equations
(2).

For type I metrics, our method can be described by the following
algorithm:
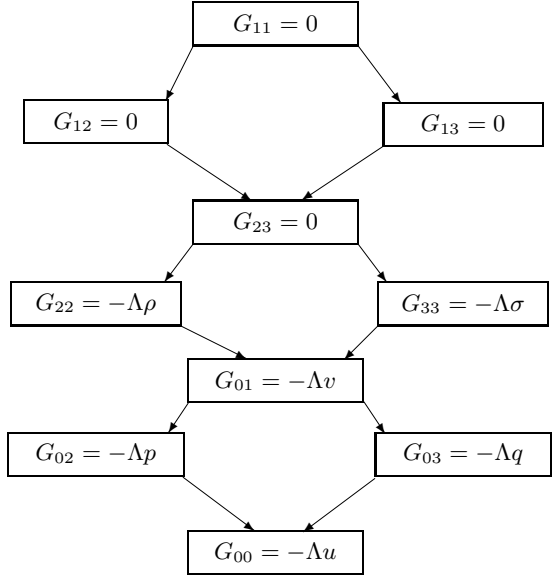
\begin{figure}[H]
    \begin{center}
\begin{picture}(204,220)
\thinlines \drawframebox{99.0}{210.0}{62.0}{16.0}{$G_{11}=0$}
\drawframebox{31.0}{173.0}{54.0}{16.0}{$G_{12}=0$}
\drawframebox{170.0}{172.0}{60.0}{16.0}{$G_{13}=0$}
\drawframebox{99.0}{135.0}{62.0}{16.0}{$G_{23}=0$}
\drawframebox{31.0}{104.0}{64.0}{16.0}{$G_{22}=-\Lambda \rho$}
\drawframebox{170.0}{104.0}{64.0}{16.0}{$G_{33}=-\Lambda\sigma$}
\drawframebox{99.0}{75.0}{66.0}{16.0}{$G_{01}=-\Lambda v$}
\drawframebox{31.0}{47.0}{66.0}{16.0}{$G_{02}=-\Lambda p$}
\drawframebox{170.0}{47.0}{66.0}{16.0}{$G_{03}=-\Lambda q$}
\drawframebox{99.0}{10.0}{66.0}{16.0}{$G_{00}=-\Lambda u$}
\drawvector{68.0}{202.0}{10.5}{-1}{-2}
\drawvector{130.0}{202.0}{16.5}{3}{-4}
\drawvector{58.0}{164.5}{32.0}{3}{-2}
\drawvector{140.0}{164.0}{31.0}{-3}{-2}
\drawvector{68.0}{127.0}{11.0}{-3}{-4}
\drawvector{130.0}{127.0}{11.0}{3}{-4}
\drawvector{63.5}{95.5}{24.5}{2}{-1}
\drawvector{138.0}{96.0}{13.0}{-1}{-1}
\drawvector{66.0}{66.5}{7.5}{-2}{-3}
\drawvector{132.5}{67.0}{8.0}{2}{-3}
\drawvector{64.0}{38.5}{27.0}{4}{-3}
\drawvector{136.5}{38.0}{27.0}{-4}{-3}
\end{picture}
\caption{The algorithm to construct the exact solutions}
    \end{center}
\end{figure}

In fact, the equations (2) can be rewritten as
\begin{equation}
G_{\mu\nu}\stackrel{\triangle}{=}
R_{\mu\nu}-\frac{1}{2}\eta_{\mu\nu}R =-\Lambda \eta_{\mu\nu},
\end{equation}
where $G_{\mu\nu}$ is the Einstein tensor. Noting the special form
of type I, we have
\begin{equation}\begin{array}{lll}
G_{11} & = & {\displaystyle
-\frac{1}{2}\left\{\frac{v_x}{v}\left(\frac{\rho_x}{\rho}+\frac{\sigma_x}{\sigma}\right)+\right.}\vspace{2mm}\\
& &  {\displaystyle\left.
\frac{1}{2}\left[\left(\frac{\rho_x}{\rho}\right)^2+\left(\frac{\sigma_x}{\sigma}\right)^2\right]-\left(
\frac{\rho_{xx}}{\rho}+\frac{\sigma_{xx}}{\sigma}\right)\right\}.}\end{array}
\end{equation}
One of the equations (15) reads
\begin{equation}
G_{11}=-\Lambda \eta_{11}.\end{equation} Solving the ODE (17) gives
\begin{equation}
v=v_0\exp\left\{\int\left[\frac{\rho_{xx}}{\rho}+\frac{\sigma_{xx}}{\sigma}-
\frac{1}{2}\left(\frac{\rho_x}{\rho}\right)^2-
\frac{1}{2}\left(\frac{\sigma_x}{\sigma}\right)^2\right]
\frac{\rho\sigma}{(\rho\sigma)_x}dx\right\},
\end{equation}
where $v_0=v_0(t,y,z)$ is an integral function depending on $t,\,y$
and $z$, provided that $(\rho\sigma)_x\neq 0$.

In particular, by taking the ansatz
\begin{equation}
\rho=\tilde{\rho}(t,y,z)\exp\{2f(t,x)\},~~\sigma=\tilde{\sigma}(t,y,z)\exp\{2f(t,x)\},
\end{equation}
(18) becomes
\begin{equation}
v=v_0f_xe^f.
\end{equation}

According to the algorithm shown in Fig 1, in a similar way we can
solve other equations in (2), and then we can construct many exact
solutions of (2).

\noindent{\bf Remark 1} {\em Our method can be used to solve the
Einstein's field equations with physically relevant energy-momentum
tensors, e.g., the tensor for perfect fluid:
$T_{\gamma\delta}=(\mu+p)u_{\gamma}u_{\delta}+pg_{\gamma\delta}$,
where $\mu>0$ is the density, $p$ is the pressure, $u$ stands for
the space-time velocity of the fluid with
$u_{\gamma}u^{\gamma}=-1$.}

\noindent{\bf Remark 2} {\em The method presented in this paper can
also be used to solve the Einstein's field equations in higher
space-time dimensions.}

{\em 4. Time-periodic solutions.} By the method presented in the
last section, we can construct many new exact solutions to the
vacuum Einstein's field equations
\begin{equation}
G_{\mu\nu}=0.
\end{equation}
For example, in the coordinates
$(\tau,\widetilde{r},\widetilde{\theta},\widetilde{\varphi})$,
taking the ansatz in (19) as follows
$$\widetilde{\rho}=-1,\quad \widetilde{\sigma}=-\sin^2\widetilde{\theta},\quad f=\ln
[\widetilde{r}+\tau+\varepsilon\sin\tau ],$$ one can obtain an
interesting solution of the form
\begin{equation}
\widetilde{\eta}_{\mu\nu}= \left(
\begin{array}{cccc}
\widetilde{\eta}_{00} & \widetilde{\eta}_{01} & \widetilde{\eta}_{02} & 0 \\
\widetilde{\eta}_{01} & 0 & 0 & 0  \\
\widetilde{\eta}_{02} & 0 & \widetilde{\eta}_{22} & 0 \\
0 & 0 & 0 & \widetilde{\eta}_{33}\\
\end{array}\right),
\end{equation}
where
\begin{equation}
\left\{\begin{array}{l} {\widetilde{\eta}}_{00}=1+2\varepsilon
\widetilde{\Omega}^+\sin\widetilde{\theta}\cos{\tau}
+\displaystyle\frac{2\widetilde{\Omega}^+\sin\widetilde{\theta}}{m}
\widetilde{\Omega},\\
\widetilde{\eta}_{01}=\displaystyle\frac{\widetilde{\Omega}^+
\sin\widetilde{\theta}}{m}\widetilde{\Omega},\\
\widetilde{\eta}_{02}=\displaystyle\frac{1}{2}(1+2\sin\widetilde{\theta})\widetilde{\Omega}^-
\left[\widetilde{r}+\tau+\varepsilon\sin\tau\right],\\
\widetilde{\eta}_{22}=-\left[\widetilde{r}+\tau+\varepsilon\sin\tau\right]^2,\\
\widetilde{\eta}_{33}=-\left[\widetilde{r}+\tau+\varepsilon\sin\tau\right]^2\sin^2\widetilde{\theta},
\end{array}\right.
\end{equation}
in which $\varepsilon$ is a parameter, $m\neq 0$ is a constant, and
\begin{equation}
\widetilde{\Omega} = exp\left(\frac{\widetilde{r}+\tau}{m}\right)+m
,\quad
\widetilde{\Omega}^{\pm}=|\tan\widetilde{\theta}/2|^{\frac12}\pm
|\tan\widetilde{\theta}/2|^{-\frac12}.
\end{equation}
The solution (22) belongs to the class of type I, also belongs to
the class of type I$\!$I$\!$I.

Making the transformation
\begin{equation}
\left\{\begin{array}{l}
t=\tau+\widetilde{r},\\
r=m+\exp{\left(\displaystyle\frac{\tau+\widetilde{r}}{m}\right)},\\
\theta=\widetilde{\theta},\\
\varphi=\widetilde{\varphi}.
\end{array}\right.
\end{equation}
Then the solution to (21) becomes, in the coordinates
$(t,r,\theta,\varphi)$,
\begin{equation}
ds^2=(dt,dr,d\theta,d\varphi)(\eta_{\mu\nu})(dt,dr,d\theta,d\varphi)^T,
\end{equation}
where
\begin{equation}
(\eta_{\mu\nu})= \left(
\begin{array}{cccc}
G & -G+\displaystyle\frac{Mr}{r-m} & QK & 0  \\
-G+\displaystyle\frac{Mr}{r-m} & G-\displaystyle\frac{2Mr}{r-m} & -QK & 0  \\
QK & -QK & -K^2 & 0 \\
0 & 0 & 0 & -K^2\sin^2\theta \\
\end{array}\right),
\end{equation}
in which
\begin{equation}
\left\{\begin{array}{l}
G=1+2\varepsilon \Omega^+\sin\theta\cos{(t-r)},\\
K=r+m\ln{|r-m|}+\varepsilon\sin{(t-r)},\\
M=\Omega^+\sin\theta,\\
Q=-\displaystyle\frac{1}{2}(1+2\sin\theta)\Omega^-.
\end{array}\right.
\end{equation}
In (28), $\Omega^{\pm}$ are defined by
\begin{equation}
\Omega^{\pm}=|\tan\theta/2|^{\frac12}\pm|\tan\theta/2|^{-\frac12}.
\end{equation}

An important property of the space-time described by (26) is given
by the following theorem.

\noindent{\bf Theorem 2} {\em When $\varepsilon$ takes its value in
the interval $(-\frac 18, \frac 18)$, i.e., $\varepsilon\in (-\frac
18, \frac 18)$, the solution (26) to the vacuum Einstein's field
equations is time-periodic.}

\noindent{\bf Proof.} Noting $\varepsilon\in (-\frac 18, \frac 18)$,
we have
\begin{equation}
\begin{array}{lll} \eta_{00} & = & {\displaystyle G=1+2\varepsilon
\Omega^+\sin\theta\cos{(t-r)}}\vspace{3mm}\\
& = & {\displaystyle
1+4\varepsilon\left(\sqrt{\left|\frac{\sin\frac{\theta}{2}}{\cos\frac{\theta}{2}}\right|}+
\sqrt{\left|\frac{\sin\frac{\theta}{2}}{\cos\frac{\theta}{2}}\right|}
\right)\times}\vspace{3mm}\\
&  & {\displaystyle
\sin\frac{\theta}{2}\cos\frac{\theta}{2}\cos(t-r)}\vspace{3mm}\\
& \ge & {\displaystyle
1-4|\varepsilon|\left(\sqrt{\left|\sin^3\frac{\theta}{2}\right|\left|\cos\frac{\theta}{2}\right|}+
\sqrt{\left|\cos^3\frac{\theta}{2}\right|\left|\sin\frac{\theta}{2}\right|}
\right)\times}\vspace{3mm}\\
&  & {\displaystyle
\left|\cos(t-r)\right|}\vspace{3mm}\\
& \ge & {\displaystyle 1-8|\varepsilon|>1-8\times\frac{1}{8}=0.}
\end{array}
\end{equation}

On the other hand, by calculations we have
\begin{equation}
\left|\begin{array}{cccc}
G & -G+\displaystyle\frac{Mr}{r-m} \vspace{3mm} \\
-G+\displaystyle\frac{Mr}{r-m} & G-\displaystyle\frac{2Mr}{r-m}
\end{array}\right|=-\frac{(Mr)^2}{(r-m)^2}<0
\end{equation}
for $r\neq 0, m$ and $\theta\neq 0,\pi$,
\begin{equation}
\left|\begin{array}{cccc}
G & -G+\displaystyle\frac{Mr}{r-m} & QK \vspace{3mm} \\
-G+\displaystyle\frac{Mr}{r-m} & G-\displaystyle\frac{2Mr}{r-m} & -QK  \vspace{3mm} \\
QK & -QK & -K^2
\end{array}\right|=K^2\frac{(Mr)^2}{(r-m)^2}>0
\end{equation}
and
\begin{equation}\begin{array}{l}
{\displaystyle \left|\begin{array}{cccc}
G & -G+\displaystyle\frac{Mr}{r-m} & QK & 0  \\
-G+\displaystyle\frac{Mr}{r-m} & G-\displaystyle\frac{2Mr}{r-m} & -QK & 0  \\
QK & -QK & -K^2 & 0 \\
0 & 0 & 0 & -K^2\sin^2\theta \\
\end{array}\right|=
}\vspace{3mm}\\
{\displaystyle\hskip 3cm
-K^4\sin^2\theta\frac{(Mr)^2}{(r-m)^2}<0}\end{array}
\end{equation}
for $r\neq 0, m$, $\theta\neq 0,\pi$ and $K\neq 0$. In next section,
we will show that $\theta = 0,\pi$, $r = 0, m$ and $K= 0$ are not
essential singularities, in other words, they are not physical
singularities.

The above discussion implies that the variable $t$ is a time
coordinate. Therefore, it follows from (27) and (28) that the
Lorentzian metric (26) is indeed a time-periodic solution of the
vacuum Einstein's field equations. This proves Theorem 2.
$\quad\quad\quad\blacksquare$

\noindent{\bf Remark 3} {\em The time-periodic solution (26) was
first obtained by us in early 2007. The authors have presented this
solution in several conferences for the past year. As mentioned
above, according to the authors' knowledge, this is the first
time-periodic solution to the Einstein's field equations.}

\noindent{\bf Remark 4} {\em In July 2006, the first author
discussed with Dr Y.-Q. Gu about some ideas presented in this paper.
We noted that for the case of type I metric, the author of \cite{gu}
constructed some exact solutions, none of which is time-periodic,
because the variable $t$ in these solutions can not be taken as the
time coordinate.}

Direct computations give us the following property of the
time-periodic solutions.

\noindent{\bf Property 1} {\em In the geometry of the space-time
(26), it holds that
\begin{equation}
\displaystyle\frac{\partial K}{\partial t}=\frac{G-1}{2M},\quad
\frac{\partial^2 K}{\partial t^2}=\frac{1}{2M}\frac{\partial
G}{\partial t},
\end{equation}
\begin{equation}
\displaystyle\frac{\partial \Omega^+}{\partial
\theta}=\frac{\Omega^-}{2\sin \theta},\quad \frac{\partial
\Omega^-}{\partial \theta}=\frac{\Omega^+}{2\sin \theta},\quad
\frac{\partial M}{\partial \theta}=Q,
\end{equation}
\begin{equation}
\displaystyle\frac{\partial G}{\partial r}=-\frac{\partial
G}{\partial t},\quad \frac{\partial^2 G}{\partial r\partial
t}=-\frac{\partial^2 G}{\partial t^2},\quad \frac{\partial^2
G}{\partial \theta\partial r}=-\frac{\partial^2 G}{\partial
\theta\partial t},
\end{equation}
\begin{equation}
\displaystyle\frac{\partial K}{\partial t}+\frac{\partial
K}{\partial r}=\frac{r}{r-m},\quad \frac{\partial^2 K}{\partial
t\partial r}=-\frac{\partial^2 K}{\partial t^2},\quad
\frac{\partial^2 K}{\partial \theta\partial r}=-\frac{\partial^2
K}{\partial t \partial \theta},
\end{equation}
\begin{equation}
\displaystyle\frac{\partial Q}{\partial \theta}=Q\cot
\theta-\frac{3\Omega^+}{4\sin \theta},\quad Q\frac{\partial
K}{\partial t}=\frac{1}{2}\frac{\partial G}{\partial \theta},\quad
\frac{\partial^2 G}{\partial \theta^2}=2\frac{\partial K}{\partial
t}\frac{\partial Q}{\partial \theta}.
\end{equation}
\begin{equation}
\displaystyle\frac{2Q\cos\theta}{\Omega^+}-\frac{3}{4}-\frac{Q^2}{(\Omega^+)^2}-\frac{1}{(\Omega^+)^2}=-\sin^2
\theta.
\end{equation}
\begin{equation}
\displaystyle 2Q\cot
\theta-\frac{3}{4}\frac{\Omega^+}{\sin\theta}-\frac{1+Q^2}{M}=-M.
\end{equation}}

The relations given above will play an important role in the future
study on the geometry of the time-periodic space-time (26).

There are several important questions which deserve further study:
(a) what is the topological structure of the time-periodic
space-time (26)? (b) does the space-time (26) have a compact Cauchy
surface? (c) an important point is to consider the structure of the
maximal globally hyperbolic part of the space-time (26), the
question is {\it whether this also exhibits time periodicity}.
Problems (b) and (c) were suggested by Andersson \cite{a}.

{\em 5. Singularity behaviors and physical properties.} This section
is devoted to the analysis of singularities and the physical
properties of the time-periodic solution (26).

For the metric (27), by a direct calculation, we have
\begin{equation}
g\stackrel{\triangle}{=}\det(\eta_{\mu\nu})=-M^2K^4\frac{r^2}{(r-m)^2}\sin^2\theta.
\end{equation}
Combing (28) and (41) gives
\begin{equation}
g=-(\Omega^+)^2\sin^4\theta
\left[r+m\ln{|r-m|}+\varepsilon\sin{(t-r)}\right]^4\displaystyle\frac{r^2}{(r-m)^2}.
\end{equation}
Noting (24), we obtain from (42) that
\begin{equation}
\begin{array}{l}  {\displaystyle
\mathcal{S}_{r=0,m}\stackrel{\triangle}{=}
\{(t,r,\theta,\varphi)|\;r=0,\;m\},} \vspace{2mm}\\
{\displaystyle \mathcal{S}_{K=0}\stackrel{\triangle}{=}
\{(t,r,\theta,\varphi)|\;
r+m\ln{|r-m|}+\varepsilon\sin{(t-r)}=0\},} \vspace{2mm}\\
{\displaystyle \mathcal{S}_{\theta=0,\pi}}\stackrel{\triangle}{=}
\{(t,r,\theta,\varphi)|\;\theta=0,\;\pi\} \end{array}
\end{equation}
are singularities for the solution metric (26).


By a direct calculation, we have
\begin{equation}
R_{\alpha\beta\gamma\delta}=0\quad {\rm and}\quad
R^{\alpha\beta\gamma\delta}=0
\quad(\alpha,\beta,\gamma,\delta=0,1,2,3).
\end{equation}
This gives
\begin{equation}
R_{\mu\nu}=0\quad (\mu,\nu=0,1,2,3)
\end{equation}
and
\begin{equation}
\|{\bf R}\| \stackrel{\triangle}{=}
R^{\alpha\beta\gamma\delta}R_{\alpha\beta\gamma\delta}=0.
\end{equation}
(45) implies that the Lorentzian metric (26) is indeed a solution of
the vacuum Einstein's field equation (21), and (46) implies that
this solution does not have any essential singularity.

According to the definition of event horizon (see e.g., Wald
\cite{wald}), it is easy to show that $\mathcal{S}_{\theta=0,\pi}$
and $\mathcal{S}_{K=0}$ are the event horizons of the space-time
(26). If $(t,r,\theta,\varphi)$ are the polar coordinates, then the
$\mathcal{S}_{r=0}$ can be regarded as a degenerate event horizon.
However, $\mathcal{S}_{r=m}$ is a new kind of singularity which is
neither event horizon nor black hole. This differs from that in
Schwartzschild space-time, in which $r=0$ corresponds to the black
hole and $r=m$ corresponds to the event horizon. In other words, the
solution (26) describes an essentially regular space-time, it does
not contain any essential singularity like black hole. It is an
interesting topic to see how to cancel these kinds of singularities
by making some coordinate transformation. Therefore, we have

\noindent{\bf Property 2}  {\em The Lorentzian metric (26) describes
a regular space-time, this space-time is Riemannian flat in the
sense of (44), it does not contain any essential singularity.
However it contains some non-essential singularities which
correspond to event horizons and some other new physical phenomena.}

\noindent{\bf Property 3}  {\em The non-essential singularities of
the space-time (26) consist of three parts $\mathcal{S}_{r=0,m}$,
$\mathcal{S}_{K=0}$ and $\mathcal{S}_{\theta=0,\pi}$.
$\mathcal{S}_{\theta=0,\pi}$ are two steady event horizons; and
$\mathcal{S}_{K=0}$ is the ``time-periodic" event horizon;
$\mathcal{S}_{r=0,m}$ are two new kinds of singularities which are
neither event horizons nor black holes. In particular, if
$(t,r,\theta,\varphi)$ are the polar coordinates, then
$\mathcal{S}_{r=0}$ can be regarded as a degenerate event horizon.}

According to the authors' knowledge, the degenerate event horizon
and the time-periodic event horizon are two new phenomena in the
space-time geometry.

\vskip 4mm

We next consider the time-periodic event horizon in detail. Without
loss of generality, we may assume that $\varepsilon$ and $m$ are
positive constants. We have two cases: $0\le r\le m$ and $r>m$.

\vskip 4mm

\noindent {\bf Case I:} $\;\; 0\le r\le m.$

Let
\begin{equation}
f(r;t)=r+m\ln{|m-r|}+\varepsilon\sin{(t-r)}.
\end{equation}
For any fixed $t\in\mathbb{R}$, it holds that
\begin{equation}
f(r;t)\longrightarrow -\infty\quad {\rm as}\;\; r\rightarrow m
\end{equation}
and
\begin{equation}
f(r;t)\longrightarrow \infty\quad {\rm as}\;\; r\rightarrow \infty.
\end{equation}
At $r=0$, we consider
\begin{equation}
m\ln{m}+\varepsilon\sin{t}=0,
\end{equation}
i.e.,
\begin{equation}
\sin{t}=-\frac{m\ln{m}}{\varepsilon}.
\end{equation}
It is obvious that (51) has a solution if and only if
\begin{equation}
\left|\frac{m\ln{m}}{\varepsilon}\right|\le 1.
\end{equation}

In what follows, we always assume condition (52). Therefore, it
follows from (51) that
\begin{equation}
f(0;t_k)=0,
\end{equation}
where
\begin{equation}
t_k=2k\pi+\arcsin\left\{-\frac{m\ln{m}}{\varepsilon}\right\},
\end{equation}
in which $k\in \mathbb{Z}$.

We now divide the discussion into two cases.

{\bf Case I-1:} $\;\; 0< m\le 1.$

In this case, we have
\begin{equation}
-\frac{m\ln{m}}{\varepsilon}\ge 0,
\end{equation}
and then
\begin{equation}
\arcsin\left\{-\frac{m\ln{m}}{\varepsilon}\right\}\ge 0.
\end{equation}
Therefore,
\begin{equation}
t_k=2k\pi+\arcsin\left\{-\frac{m\ln{m}}{\varepsilon}\right\}\quad
(k=0,1,2\cdots).
\end{equation}

{\bf Case I-2:} $\;\; 1< m.$

In this case,
\begin{equation}
-\frac{m\ln{m}}{\varepsilon}< 0,
\end{equation}
and
\begin{equation}
\arcsin\left\{-\frac{m\ln{m}}{\varepsilon}\right\}<0.
\end{equation}
Thus, we shall take
\begin{equation}
t_k=2(k+1)\pi+\arcsin\left\{-\frac{m\ln{m}}{\varepsilon}\right\}\quad
(k=0,1,2\cdots).
\end{equation}

In both case I-1 and case I-2, by fixing $k\in\{0,1,2,\cdots\}$ and
noting (48), we see that there exists a maximum $r_-\in [0,m)$ such
that, for any given $r\in [0,r_-]$, the equation for $t$
\begin{equation}
f(r;t)=0
\end{equation}
has solutions. When $r=r_-$, we denote the solution by $t_k^-$. It
holds that
\begin{equation}
f(r_-;t_k^-)=0.
\end{equation}
Summarizing the above discussion, we observe that, for case I, the
time-periodic event horizons are given in Fig. 2.
\begin{figure}[H]
    \begin{center}
\begin{picture}(188,168)
\thinlines \drawvector{22.0}{12.0}{152.0}{0}{1}
\drawvector{22.0}{12.0}{160.0}{1}{0}
\drawdashline{130.0}{12.0}{130.0}{152.0}
\drawdashline{90.0}{12.0}{90.0}{156.0}
\path(22.0,30.0)(22.0,30.0)(22.88,30.46)(23.75,30.95)(24.57,31.42)(25.39,31.88)(26.18,32.34)(26.95,32.8)(27.7,33.27)(28.43,33.72)
\path(28.43,33.72)(29.14,34.16)(29.84,34.61)(30.51,35.05)(31.15,35.5)(31.79,35.93)(32.4,36.36)(33.01,36.79)(33.59,37.2)(34.15,37.62)
\path(34.15,37.62)(34.72,38.04)(35.26,38.47)(35.79,38.87)(36.29,39.27)(36.79,39.68)(37.29,40.08)(37.77,40.47)(38.25,40.86)(38.7,41.25)
\path(38.7,41.25)(39.15,41.63)(39.59,42.02)(40.02,42.4)(40.45,42.77)(40.87,43.15)(41.29,43.5)(41.7,43.86)(42.12,44.22)(42.52,44.59)
\path(42.52,44.59)(42.91,44.93)(43.3,45.29)(43.7,45.63)(44.09,45.97)(44.48,46.31)(44.87,46.65)(45.27,46.97)(45.66,47.31)(46.05,47.63)
\path(46.05,47.63)(46.45,47.95)(46.86,48.27)(47.26,48.58)(47.66,48.88)(48.08,49.18)(48.48,49.49)(48.91,49.79)(49.34,50.09)(49.77,50.38)
\path(49.77,50.38)(50.22,50.66)(50.68,50.95)(51.15,51.22)(51.61,51.5)(52.09,51.77)(52.59,52.04)(53.09,52.31)(53.61,52.58)(54.15,52.84)
\path(54.15,52.84)(54.68,53.09)(55.25,53.34)(55.81,53.59)(56.4,53.83)(57.0,54.06)(57.63,54.31)(58.27,54.54)(58.93,54.77)(59.59,55.0)
\path(59.59,55.0)(60.29,55.22)(61.0,55.43)(61.74,55.65)(62.49,55.86)(63.27,56.08)(64.06,56.27)(64.88,56.47)(65.72,56.68)(66.59,56.86)
\path(66.59,56.86)(67.5,57.06)(68.41,57.25)(69.36,57.43)(70.33,57.61)(71.33,57.79)(72.36,57.95)(73.41,58.13)(74.49,58.29)(75.61,58.45)
\path(75.61,58.45)(76.75,58.61)(77.93,58.77)(79.13,58.91)(80.36,59.06)(81.63,59.2)(82.94,59.34)(84.29,59.49)(85.66,59.61)(87.06,59.75)
\path(87.06,59.75)(88.51,59.86)(89.98,59.99)(90.0,60.0)
\path(22.0,104.0)(22.0,104.0)(22.82,104.41)(23.63,104.83)(24.44,105.26)(25.21,105.69)(25.98,106.12)(26.75,106.55)(27.48,106.98)(28.2,107.41)
\path(28.2,107.41)(28.92,107.84)(29.61,108.29)(30.31,108.73)(30.97,109.16)(31.63,109.61)(32.29,110.04)(32.93,110.48)(33.56,110.91)(34.18,111.36)
\path(34.18,111.36)(34.79,111.8)(35.38,112.23)(35.97,112.68)(36.55,113.12)(37.12,113.55)(37.7,114.0)(38.26,114.43)(38.8,114.87)(39.34,115.3)
\path(39.34,115.3)(39.88,115.73)(40.41,116.16)(40.95,116.59)(41.47,117.02)(41.98,117.44)(42.5,117.87)(43.01,118.3)(43.52,118.72)(44.02,119.13)
\path(44.02,119.13)(44.52,119.55)(45.02,119.95)(45.52,120.37)(46.01,120.77)(46.51,121.18)(47.0,121.58)(47.5,121.97)(47.98,122.37)(48.48,122.75)
\path(48.48,122.75)(48.97,123.13)(49.47,123.51)(49.97,123.88)(50.47,124.26)(50.99,124.62)(51.49,125.0)(52.0,125.34)(52.52,125.7)(53.04,126.05)
\path(53.04,126.05)(53.56,126.4)(54.09,126.73)(54.63,127.05)(55.18,127.38)(55.72,127.7)(56.27,128.01)(56.84,128.33)(57.4,128.63)(57.99,128.92)
\path(57.99,128.92)(58.58,129.22)(59.16,129.5)(59.77,129.76)(60.38,130.04)(61.02,130.3)(61.65,130.55)(62.29,130.8)(62.95,131.04)(63.63,131.26)
\path(63.63,131.26)(64.3,131.5)(65.0,131.7)(65.7,131.91)(66.43,132.11)(67.16,132.3)(67.91,132.48)(68.68,132.66)(69.45,132.83)(70.25,132.98)
\path(70.25,132.98)(71.05,133.13)(71.88,133.26)(72.72,133.39)(73.58,133.51)(74.47,133.63)(75.36,133.72)(76.27,133.8)(77.2,133.88)(78.16,133.95)
\path(78.16,133.95)(79.13,134.01)(80.11,134.07)(81.13,134.1)(82.16,134.13)(83.2,134.13)(84.29,134.14)(85.37,134.13)(86.5,134.11)(87.63,134.08)
\path(87.63,134.08)(88.8,134.05)(89.98,134.0)(90.0,134.0)
\drawdashline{90.0}{134.0}{22.0}{134.0}
\drawdashline{90.0}{60.0}{22.0}{60.0}
\drawcenteredtext{13.0}{30.0}{$t_0$}
\drawcenteredtext{13.0}{104.0}{$t_k$}
\drawcenteredtext{13.0}{60.0}{$t_0^-$}
\drawcenteredtext{13.0}{134.0}{$t_k^-$}
\drawcenteredtext{13.0}{162.0}{$t$}
\drawcenteredtext{180.0}{2.0}{$r$}
\drawcenteredtext{130.0}{2.0}{$m$}
\drawcenteredtext{90.0}{2.0}{$r_-$}
\drawcenteredtext{16.0}{4.0}{$0$}
\end{picture}
\caption{Time-periodic event horizons for case I}
    \end{center}
\end{figure}
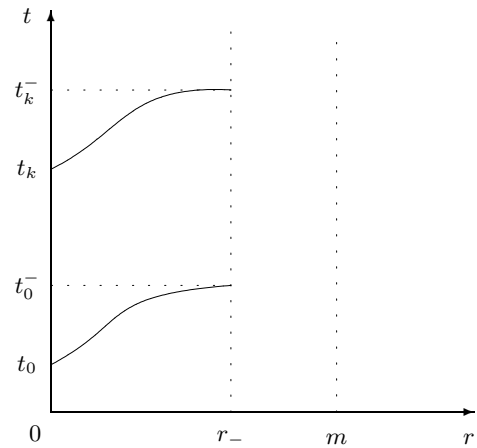

\noindent {\bf Case I$\!$I:} $\;\; r> m.$

Similar to the discussion of case I, in this case the time-periodic
event horizons are given in Fig. 3.
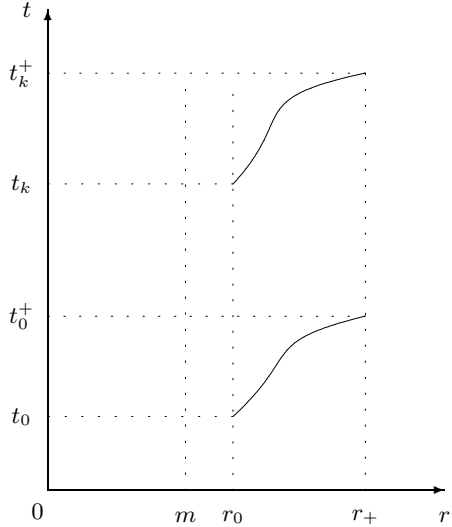
\begin{figure}[H]
    \begin{center}
\begin{picture}(178,180)
\thinlines \drawvector{20.0}{12.0}{182.0}{0}{1}
\drawvector{20.0}{12.0}{150.0}{1}{0}
\drawdashline{72.0}{12.0}{72.0}{164.0}
\drawdashline{90.0}{12.0}{90.0}{162.0}
\drawdashline{140.0}{12.0}{140.0}{170.0}
\path(90.0,40.0)(90.0,40.0)(90.75,40.7)(91.5,41.4)(92.22,42.09)(92.91,42.77)(93.58,43.45)(94.24,44.09)(94.86,44.75)(95.47,45.38)
\path(95.47,45.38)(96.07,46.0)(96.63,46.61)(97.17,47.22)(97.71,47.81)(98.22,48.4)(98.71,48.97)(99.19,49.52)(99.64,50.08)(100.1,50.62)
\path(100.1,50.62)(100.52,51.16)(100.94,51.69)(101.33,52.19)(101.72,52.7)(102.1,53.2)(102.46,53.69)(102.8,54.18)(103.14,54.65)(103.47,55.12)
\path(103.47,55.12)(103.8,55.56)(104.11,56.01)(104.41,56.45)(104.69,56.9)(104.99,57.31)(105.25,57.73)(105.53,58.16)(105.8,58.55)(106.05,58.95)
\path(106.05,58.95)(106.32,59.34)(106.57,59.73)(106.82,60.12)(107.05,60.48)(107.3,60.86)(107.55,61.22)(107.78,61.58)(108.02,61.93)(108.27,62.26)
\path(108.27,62.26)(108.5,62.61)(108.75,62.94)(109.0,63.27)(109.24,63.61)(109.49,63.93)(109.74,64.23)(110.0,64.55)(110.25,64.87)(110.52,65.16)
\path(110.52,65.16)(110.8,65.47)(111.08,65.76)(111.36,66.06)(111.66,66.36)(111.97,66.63)(112.27,66.93)(112.6,67.2)(112.92,67.48)(113.27,67.76)
\path(113.27,67.76)(113.63,68.04)(114.0,68.3)(114.36,68.58)(114.75,68.84)(115.16,69.12)(115.58,69.38)(116.02,69.65)(116.47,69.91)(116.92,70.16)
\path(116.92,70.16)(117.41,70.44)(117.91,70.69)(118.42,70.94)(118.96,71.2)(119.52,71.47)(120.08,71.73)(120.67,71.98)(121.3,72.25)(121.92,72.51)
\path(121.92,72.51)(122.58,72.76)(123.27,73.02)(123.97,73.29)(124.69,73.55)(125.44,73.8)(126.22,74.08)(127.02,74.33)(127.83,74.61)(128.69,74.87)
\path(128.69,74.87)(129.58,75.15)(130.47,75.41)(131.41,75.69)(132.38,75.97)(133.38,76.25)(134.39,76.54)(135.46,76.81)(136.53,77.11)(137.66,77.4)
\path(137.66,77.4)(138.8,77.69)(139.99,77.98)(140.0,78.0)
\path(90.0,128.0)(90.0,128.0)(90.69,128.77)(91.38,129.52)(92.03,130.27)(92.66,131.02)(93.27,131.75)(93.86,132.46)(94.42,133.16)(94.97,133.86)
\path(94.97,133.86)(95.49,134.55)(95.99,135.22)(96.46,135.88)(96.91,136.52)(97.36,137.16)(97.78,137.8)(98.19,138.42)(98.58,139.03)(98.96,139.63)
\path(98.96,139.63)(99.32,140.22)(99.66,140.8)(100.0,141.38)(100.3,141.94)(100.61,142.5)(100.91,143.05)(101.19,143.58)(101.46,144.11)(101.72,144.63)
\path(101.72,144.63)(101.97,145.14)(102.22,145.66)(102.46,146.14)(102.69,146.63)(102.91,147.11)(103.11,147.6)(103.33,148.05)(103.53,148.52)(103.74,148.97)
\path(103.74,148.97)(103.94,149.41)(104.13,149.86)(104.33,150.28)(104.52,150.71)(104.71,151.13)(104.91,151.53)(105.1,151.94)(105.28,152.35)(105.49,152.74)
\path(105.49,152.74)(105.69,153.13)(105.88,153.52)(106.1,153.88)(106.3,154.27)(106.52,154.63)(106.74,155.0)(106.97,155.35)(107.21,155.71)(107.44,156.05)
\path(107.44,156.05)(107.69,156.39)(107.97,156.74)(108.24,157.07)(108.52,157.41)(108.8,157.72)(109.11,158.05)(109.42,158.38)(109.77,158.69)(110.11,159.0)
\path(110.11,159.0)(110.47,159.32)(110.83,159.63)(111.22,159.94)(111.63,160.24)(112.05,160.53)(112.5,160.83)(112.97,161.13)(113.44,161.42)(113.94,161.72)
\path(113.94,161.72)(114.46,162.0)(115.0,162.3)(115.57,162.58)(116.14,162.86)(116.75,163.14)(117.38,163.44)(118.05,163.72)(118.72,164.0)(119.42,164.27)
\path(119.42,164.27)(120.16,164.55)(120.92,164.83)(121.72,165.11)(122.52,165.39)(123.38,165.67)(124.25,165.96)(125.16,166.25)(126.1,166.52)(127.05,166.8)
\path(127.05,166.8)(128.05,167.08)(129.1,167.36)(130.16,167.66)(131.27,167.94)(132.39,168.22)(133.58,168.52)(134.77,168.8)(136.02,169.11)(137.3,169.39)
\path(137.3,169.39)(138.63,169.69)(139.99,170.0)(140.0,170.0)
\drawdashline{140.0}{170.0}{20.0}{170.0}
\drawdashline{90.0}{128.0}{20.0}{128.0}
\drawdashline{90.0}{40.0}{20.0}{40.0}
\drawdashline{20.0}{78.0}{140.0}{78.0}
\drawcenteredtext{12.0}{194.0}{$t$}
\drawcenteredtext{10.0}{170.0}{$t_k^+$}
\drawcenteredtext{10.0}{128.0}{$t_k$}
\drawcenteredtext{10.0}{40.0}{$t_0$}
\drawcenteredtext{10.0}{78.0}{$t_0^+$}
\drawcenteredtext{72.0}{2.0}{$m$}
\drawcenteredtext{90.0}{2.0}{$r_0$}
\drawcenteredtext{140.0}{2.0}{$r_+$}
\drawcenteredtext{170.0}{2.0}{$r$} \drawcenteredtext{16.0}{4.0}{$0$}
\end{picture}
\caption{Time-periodic event horizons for case I$\!$I}
    \end{center}
\end{figure}
In Fig. 3, $r_0$ and $r_+$ are defined in the following way: noting
(48) and (49), we see that there exists a minimum $r_0\in
(m,\infty)$ and a maximum $r_+\in (m,\infty)$ such that, for any
given $r\in [r_0,r_+]$, the equation for $t$
\begin{equation}
f(r;t)=0
\end{equation}
has solutions. In particular, when $r=r_0$ (resp. $r=r_+$), we
denote the solution by $t_k$ (resp. by $t_k^+$). That is to say, it
holds that
\begin{equation}
f(r_0;t_k)=0\quad {\rm and}\quad f(r_+;t_k^+)=0.
\end{equation}
Therefore, we have proved the following property.

\noindent{\bf Property 4}  {\em The non-essential singularities of
the space-time (26) consists of three parts $r=0$, $r=m$ and
$r+m\ln{|r-m|}+\varepsilon\sin{(t-r)}=0$. $r=0$ is a degenerate
event horizon, $r=m$ is a steady event horizon, and
$r+m\ln{|r-m|}+\varepsilon\sin{(t-r)}=0$ are the ``time-periodic"
event horizons. Time-periodic event horizons form and disappear in
finite times, they propagate time-periodically.}

On the other hand, by some elementary matrix transformations, the
metric $(\eta_{\mu\nu})$ can be reduced to
\begin{equation}
(\hat{\eta}_{\mu\nu})={\rm diag}\left\{G,-\frac{M^2r^2}{G(r-m)^2},
-K^2,-K^2\sin^2\theta\right\}.
\end{equation}
Noting that (28) gives
\begin{equation}\begin{array}{l}
(\hat{\eta}_{\mu\nu})  \sim  {\rm diag}\left\{1+2\varepsilon
\Omega^+\sin\theta\cos{(t-r)},\right.\vspace{2mm}\\
\hskip 1.5cm {\displaystyle \left.
-\frac{(\Omega^+)^2\sin^2\theta}{1+2\varepsilon
\Omega^+\sin\theta\cos{(t-r)}},
-r^2,-r^2\sin^2\theta\right\}}.\end{array}
\end{equation}
In (66), we have made use of the fact that, when $r$ is large
enough, it holds that $K\sim r$ because of the second equation in
(28). (66) implies that the space-time (26) is not homogenous and
not asymptotically flat, more precisely not asymptotically
Minkowski, because the first two components in (66) depend strongly
on the angle $\theta$. Therefore, we have

\noindent{\bf Property 5}  {\em The space-time (26) is not
homogenous and not asymptotically flat.}

\noindent{\bf Remark 5} {\em Property 5 perhaps has some new
applications in cosmology due to the recent WMAP data, since the
recent WMAP data show that our Universe exists anisotropy (see
\cite{wmap}). This inhomogenous property of the new space-time (26)
may provide a way to give an explanation of this phenomena.}

\vskip 3mm

Summarizing the above discussion gives the following theorem.

\noindent{\bf Theorem 3} {\em The vacuum Einstein's field equations
have a time-periodic solution (26), this solution describes a
regular space-time, which has vanishing Riemann curvature tensor but
is not homogenous and not asymptotically flat. This space-time does
not contain any essential singularity, but contains some
non-essential singularities which correspond to steady event
horizons, time-periodic event horizon and some other new physical
phenomena.}

{\em 6. Summary and discussion.} In this paper we describe a new
method to find exact solutions to the Einstein's field equations
(1). Using our method, we can construct some important exact
solutions including the time-periodic solutions of the vacuum
Einstein's field equations. We also analyze the singularities of the
time-periodic solutions and investigate some new physical phenomena
enjoyed by these new space-times.

We remark that, by using our method, we can obtain almost all known
solutions to the Einstein's field equations. Our method can also be
used to find exact solutions of the higher dimensional Einstein's
field equations, which play an important role in string theory. The
structures of these new space-times, the behaviors of their
singularities and some new nonlinear phenomena appeared in the
time-periodic solutions are very interesting and important. We
expect some applications of these new phenomena and the
time-periodic solutions in modern cosmology and general relativity.

The authors thank Professors Lars Andersson, Chong-Ming Xu,
Shing-Tung Yau, Dr. Wen-Rong Dai, Chun-Lei He and Fu-Wen Shu for
helpful discussions and valuable suggestions. It was Professor Yau
who first brought to us the open problem of finding time-periodic
solutions to the Einstein's field equations. The work of Kong was
supported in part by the NSF of China (Grant No. 10671124) and the
NCET of China (Grant No. NCET-05-0390); the work of Liu was
supported by the NSF and NSF of China.

\end{document}